\documentclass[12pt,preprint]{aastex}

\usepackage{amsmath}
\usepackage{graphicx}
\usepackage{comment}
\usepackage{color}           % For color text: \color command
\definecolor{DarkGreen}{rgb}{0.0,0.45,0.0}  % define a custom color

\newcommand{\fig}[1]{Fig.~\ref{#1}}

\newcommand{\sect}[1]{Section~\ref{#1}}

\begin{document}
\title{Downward catastrophe of solar magnetic flux ropes}
\author{Quanhao Zhang\altaffilmark{1,4}, Yuming Wang\altaffilmark{1,2}, Youqiu Hu\altaffilmark{1}, Rui Liu\altaffilmark{1,3}}
\altaffiltext{1}{CAS Key Laboratory of Geospace Environment, Department of Geophysics and Planetary Sciences, University of Science and Technology of China, Hefei 230026, China}
\altaffiltext{2}{Synergetic Innovation Center of Quantum Information \& Quantum Physics, University of Science and Technology of China, Hefei, Anhui 230026, China}
\altaffiltext{3}{Collaborative Innovation Center of Astronautical Science and Technology, China}
\altaffiltext{4}{Mengcheng National Geophysical Observatory, School of Earth and Space Sciences, University of Science and Technology of China, Hefei 230026, China}
\email{zhangqh@mail.ustc.edu.cn}

\begin{abstract}
2.5D time-dependent ideal magnetohydrodynamic (MHD) models in Cartesian coordinates were used in previous studies to seek MHD equilibria involving a magnetic flux rope embedded in a bipolar, partially open background field. As demonstrated by these studies, the equilibrium solutions of the system are separated into two branches: the flux rope sticks to the photosphere for solutions at the lower branch but is suspended in the corona for those at the upper branch. Moreover, a solution originally at the lower branch jumps to the upper, as the related control parameter increases and reaches a critical value, and the associated jump is here referred to as upward catastrophe. The present paper advances these studies in three aspects. First, the magnetic field is changed to be force-free. The system still experiences an upward catastrophe with an increase in each control parameter. Secondly, under the force-free approximation, there also exists a downward catastrophe, characterized by a jump of a solution from the upper branch to the lower. Both catastrophes are irreversible processes connecting the two branches of equilibrium solutions so as to form a cycle. Finally, the magnetic energy in the numerical domain is calculated. It is found that there exists a magnetic energy release for both catastrophes. The Amp\`{e}re's force, which vanishes everywhere for force-free fields, appears only during the catastrophes and does a positive work, which serves as a major mechanism for the energy release. The implications of the downward catastrophe and its relevance to solar activities are briefly discussed.
\end{abstract}


\section{Introduction}
\label{sec:introduction}
\par
Coronal magnetic flux ropes are believed to have close relationship with solar eruptive activities \citep[e.g.][]{Gibson2006a,Labrosse2010a,Chen2011a}, including prominence/filament eruptions, flares, and coronal mass ejections (CMEs), which are widely considered as different manifestations of a single physical process \citep[e.g.][]{Low1996a,Wu1999a,Torok2011a,Zhang2014a}, corresponding to a sudden destabilization of the coronal magnetic configuration \citep{Archontis2008a}. Flux ropes can be triggered to erupt by many different mechanisms such as magnetic reconnections and various instabilities \citep[e.g.][]{Antiochos1999a,Chen2000a,Moore2001a,Kliem2006a}. It was also suggested by many authors that catastrophe could be responsible for the solar eruptive activities \citep{Priest1990a,Forbes1995a,Lin2004b,Zhang2007a}. Catastrophe occurs by a catastrophic loss of equilibrium as a control parameter of the magnetic system exceeds a critical value \citep{vanTend1978a,Forbes1990a,Isenberg1993a}. Here the control parameter characterizes the physical properties of the magnetic conﬁguration. Any parameter can be selected as a control parameter provided that different values of this parameter will result in different configurations \citep{Lin2002b,Wang2003a,Su2011a}, and different kinds of control parameters correspond to different evolutionary scenarios \citep{Kliem2014a}. During catastrophe, magnetic free energy is quickly released and converted to kinetic and thermal energy \citep{Chen2007a}. Catastrophe and instability are intimately related in the evolution of different magnetic systems \citep{Kliem2014a,Longcope2014a}.
\par
Many solar eruptive activities originate from active regions \citep{Benz2008a,Chen2011a}, and Cartesian coordinates are widely used to investigate active region activities. In Cartesian coordinates, both analytical and numerical analyses have been performed to explore catastrophic behaviors of flux ropes in bipolar background field. If the bipolar field is completely closed, no catastrophe occurs for the flux rope of finite cross section \citep{Hu2000a}; only if the radius of the flux rope is small compared to the length scale of the photospheric magnetic field will there exist a catastrophe \citep{Forbes1991a,Lin2002b}. During the catastrophe, the flux rope, originally attached to the photosphere, loses equilibrium at a critical value of the control parameter and reaches a new equilibrium levitating in the corona. In a partially open bipolar field, however, \cite{Hu2001a} found that catastrophe also occurs for the flux rope of finite cross section. These studies imply that there are two branches of equilibrium states of a flux rope with catastrophe: the lower branch and the upper branch. The flux rope sticks to the solar surface for solutions at the lower branch and levitates in the corona for those at the upper branch, with a vertical current sheet below it. The catastrophe mentioned above corresponds to a jump from the lower branch to the upper, and thus is called ``upward catastrophe'' hereinafter.
\par
All the previous studies only analysed the upward catastrophe. Some questions are still standing in theory: is there a catastrophe during which the flux rope falls back from the upper branch to the lower (called ``downward catastrophe'' hereinafter)? Will it also release magnetic free energy? To answer these questions, we follow the work by \cite{Hu2001a} to study the equilibrium solutions, but change to force-free fields. The motivations of using a force-free field structure rather than a magnetostatic structure are as follows. First, the strong magnetic fields over active regions are usually considered to be force-free \citep{Low1977a}. Second, under force-free conditions, the system is dominated by magnetic fields, and its energy is limited to magnetic energy so as to substantially simplify the energy analysis (see \sect{sec:energy}). By analysing the evolution of the equilibrium solutions versus the control parameters in Cartesian coordinates, the properties of the catastrophes in partially open bipolar background field under the force-free approximation are investigated. We mainly focus on the existence of the downward catastrophe, and the evolutions of the magnetic energy during the catastrophes. The sections are arranged as follows: simulation methods are introduced in \sect{sec:equation}; two kinds of catastrophes are demonstrated in \sect{sec:catastrophe}; the variations of magnetic energy during upward and downward catastrophes are analysed in \sect{sec:energy}; by summarizing the simulation results, the whole evolution of a flux rope in partially open bipolar field is illustrated in \sect{sec:profile}; the mechanism by which magnetic energy is released is analysed in \sect{sec:work}; the significance of catastrophe in both observational and theoretical analyses are discussed in \sect{sec:discuss}.
\par

\section{Basic equations and the initial and boundary conditions}
\label{sec:equation}
A Cartesian coordinate system is taken and the magnetic flux function is used here to denote the magnetic fields as follows:
\begin{align}
\textbf{B}=\triangledown\times(\psi\hat{\textbf{\emph{z}}})+B_z\hat{\textbf{\emph{z}}},\label{equ:mf}
\end{align}
and the 2.5-D MHD equations can be written in the non-dimensional form:
\begin{align}
&\frac{\partial\rho}{\partial t}+\triangledown\cdot(\rho\textbf{\emph{v}})=0,\label{equ:cal-st}\\
&\frac{\partial\textbf{\emph{v}}}{\partial t}+\textbf{\emph{v}}\cdot\triangledown\textbf{\emph{v}}+\triangledown T +\frac{T}{\rho}\triangledown\rho+\frac{2}{\rho\beta_0}(\vartriangle\psi\triangledown\psi+B_z\triangledown B_z+\triangledown\psi\times\triangledown B_z)+g\hat{\textbf{\emph{y}}}=0,\\
&\frac{\partial\psi}{\partial t}+\textbf{\emph{v}}\cdot\triangledown\psi=0,\\
&\frac{\partial B_z}{\partial t}+\triangledown\cdot(B_z\textbf{\emph{v}})+(\triangledown\psi\times\triangledown v_z)\cdot\hat{\textbf{\emph{z}}}=0,\\
&\frac{\partial T}{\partial t}+\textbf{\emph{v}}\cdot\triangledown T +(\gamma-1)T\triangledown\cdot\textbf{\emph{v}}=0,\label{equ:cal-en}
\end{align}
where $\rho, \textbf{\emph{v}}, T, \psi$ correspond to the density, velocity, temperature and magnetic flux function, respectively; $B_z$ and $v_z$ denote the z-component of the magnetic field and the velocity, parallel to the axis of the flux rope; $g$ is the normalized gravity, $\beta_0=2\mu_0\rho_0RT_0L_0^2/\psi_0^2=0.1$ is the characteristic ratio of the gas pressure to the magnetic pressure, where $\mu_0$ is the vacuum magnetic permeability, and $R$ is the gas constant; $\rho_0=3.34\times10^{-13}\mathrm{~kg~m^{-3}}$, $T_0=10^6\mathrm{~K}$, $L_0=10^7\mathrm{~m}$, and $\psi_0=3.73\times10^3\mathrm{~Wb~m^{-1}}$ are the characteristic values of density, temperature, length and magnetic flux function, respectively. The radiation and heat conduction in the energy equation have been neglected. 
\par
The initial corona is isothermal and static with
\begin{align}
T_c\equiv T(0,x,y)=1\times10^6K,\ \  \rho_c\equiv\rho(0,x,y)=\rho_0\exp^{-gy}.
\end{align}
The background field, assumed to be symmetrical relative to the $y$-axis, is a partially open bipolar magnetic field, which is calculated by complex variable method \citep[see][]{Hu2001a}. A positive and a negative surface magnetic charges are located at $y=0$ within $5\mathrm{~Mm}<x<35\mathrm{~Mm}$ and $-35\mathrm{~Mm}<x<-5\mathrm{~Mm}$, respectively. The computational domain is taken to be $0\leqslant x\leqslant300\mathrm{~Mm}$, $0\leqslant y\leqslant300\mathrm{~Mm}$, with symmetrical conditions used for the left side ($x=0$). The lower boundary $y=0$ corresponds to the photosphere. During the simulation, top ($y=300\mathrm{~Mm}$) and right ($x=300\mathrm{~Mm}$) boundary conditions are fixed. Fixed boundary conditions greatly help to maintain stability of the simulation. We have tried the simulations with computational domains of different size (e.g. $0\leqslant x\leqslant200\mathrm{~Mm}$), and found that the size of the domain does not influence the existence of the catastrophes. Although the critical values of the control parameters at which catastrophe takes place vary slightly with the domain size, the deviations are smaller than 10\%. This indicates that the computational domain used here is large enough so that the influence of the boundary condition on the evolution of the magnetic system is almost negligible. 
\par
Starting from the background field, with the same procedures as in \cite{Hu2001b}, a magnetic flux rope emerges from the central area of the base, resulting in an equilibrium state with a flux rope sticking to the photosphere, i.e. a lower-branch solution, which is the initial state of our simulation (see Fig. 1(a) in \cite{Hu2001b}).
\par
With the initial and boundary conditions, equations \ref{equ:cal-st} to \ref{equ:cal-en} are solved by the multi-step implicit scheme \citep{Hu1989a} to let the system evolve to equilibrium states. Relaxation method was used by \cite{Hu2004a} to obtain force-free equilibrium solutions in spherical coordinates. This method resets the temperature and density in the computational domain to their initial values, so that the pressure gradient force is always balanced everywhere by the gravitational force. Using the same method in Cartesian coordinates, we obtain force-free equilibrium solutions. Moreover, during the whole simulation, both numerical and physical magnetic reconnections are prohibited. In theory, $\psi$ is constant along a current sheet. Any reconnection will reduce the value of $\psi$ at the reconnection site. Therefore, by reassigning $\psi$ with the initial value at each time step along the entire current sheet (if it exists), we keep $\psi$ invariant, so that reconnections are completely prevented across the current sheet.
\par
In this paper, we select the axial magnetic flux passing through the cross section of the flux rope, $\Phi_z$, and the annular magnetic flux of the rope per unit length along $z$-direction, $\Phi_p$, which is simply the difference of $\psi$ between the rope axis and the outer border of the rope, as the control parameters. The varying $\Phi_z$ and $\Phi_p$ represent the evolutionary scenarios, e.g. flux emergence and twisting/untwisting motions. If not changed manually, $\Phi_z$ and $\Phi_p$ of the rope should be maintained to be conserved, which is achieved by special numerical measures: following similar procedures proposed by \cite{Hu2003a}, a slight adjustment of $B_z$ inside the flux rope and $\psi$ at the rope axis is taken at the end of each time step. The evolution of the equilibrium solutions is described by the geometric parameters of the flux rope, including the height of the rope axis, $H$, and the length of the current sheet below the rope, $L_c$. Note that at the lower branch, $L_c=0$, and at the upper branch, $L_c$ is finite.

\section{Simulation results}

\subsection{Upward and downward catastrophes}
\label{sec:catastrophe}
\par
Starting from the initial state, equilibrium solutions with different control parameters are calculated in the following way: from 0 to $100 \tau_A$, $\Phi_z$ and $\Phi_p$ are linearly changed from the same initial values to target values ($\Phi_z^1$, $\Phi_p^1$), and from $100 \tau_A$ to $200 \tau_A$, the two fluxes are kept invarant till the magnetic system reaches an equilibrium state. Here $\tau_A=L_0^2\sqrt{\mu_0\rho_0}/\psi_0 =17.4\mathrm{~s}$ is the typical Alfv\'{e}n transit time. The final state of this calculation at $t=200 \tau_A$ is regarded as the equilibrium solution with ($\Phi_z^1$, $\Phi_p^1$). Similar calculations are repeated for different target values of $\Phi_z$ and $\Phi_p$ to obtain different equilibrium solutions.
%Similar calculations are repeated for different target values of $\Phi_z$ but with the same $\Phi_p=14.9\times10^{3}\mathrm{~Wb~m}^{-1}$.
\par
\begin{figure*}
\includegraphics[width=\hsize]{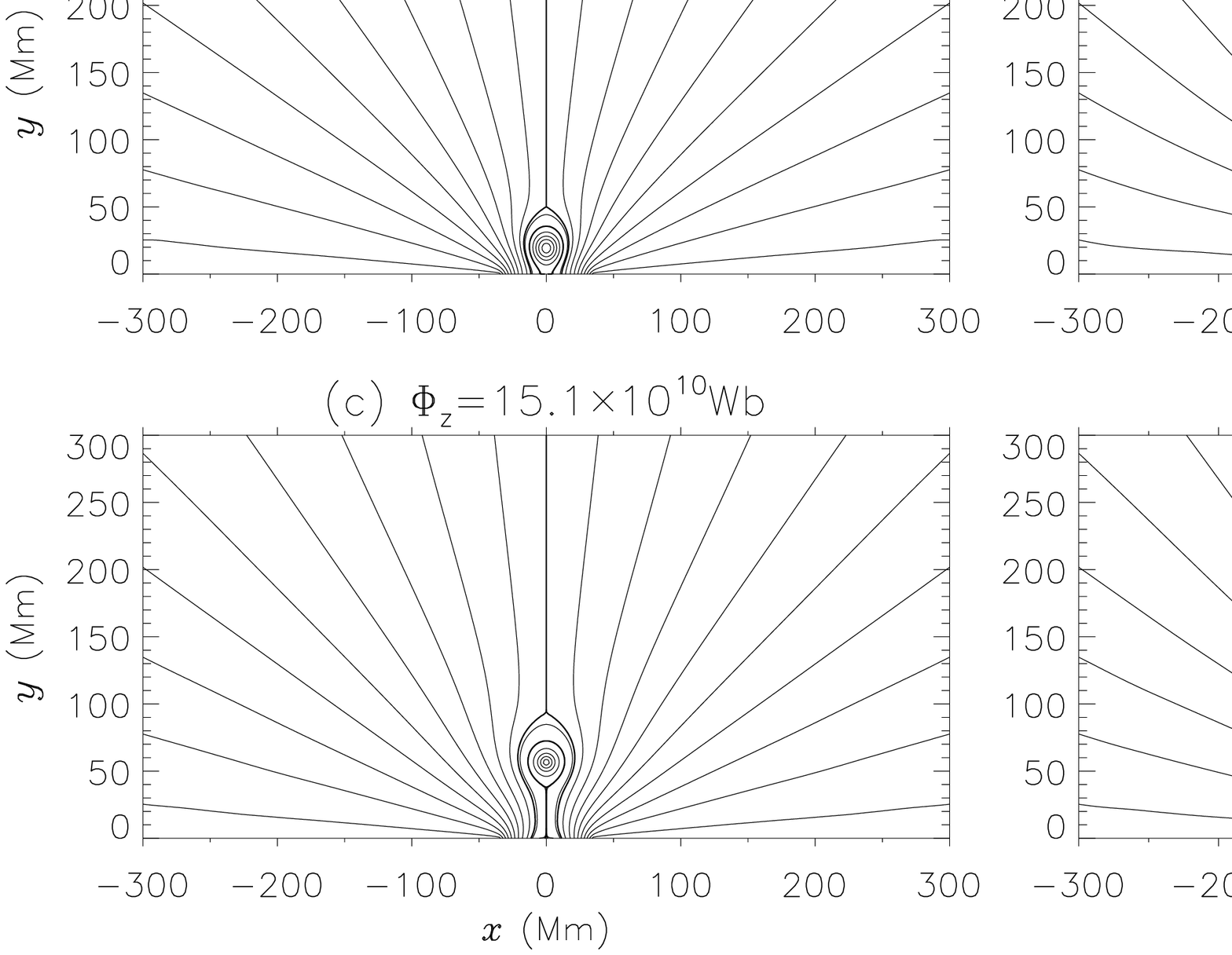}
\caption{Magnetic configurations of the flux rope system (a) right before and (b) after the upward catastrophic point, (c) right before and (d) after the downward catastrophic point. The annular flux is fixed to be $\Phi_p = 14.9\times 10^3\mathrm{~Wb\cdot m^{-1}}$ for the system, whereas the axial flux $\Phi_z$ is taken as the
control parameter.}\label{fig:conf}
\end{figure*}
\begin{figure*}
\centerline{\includegraphics[width=0.8\textwidth]{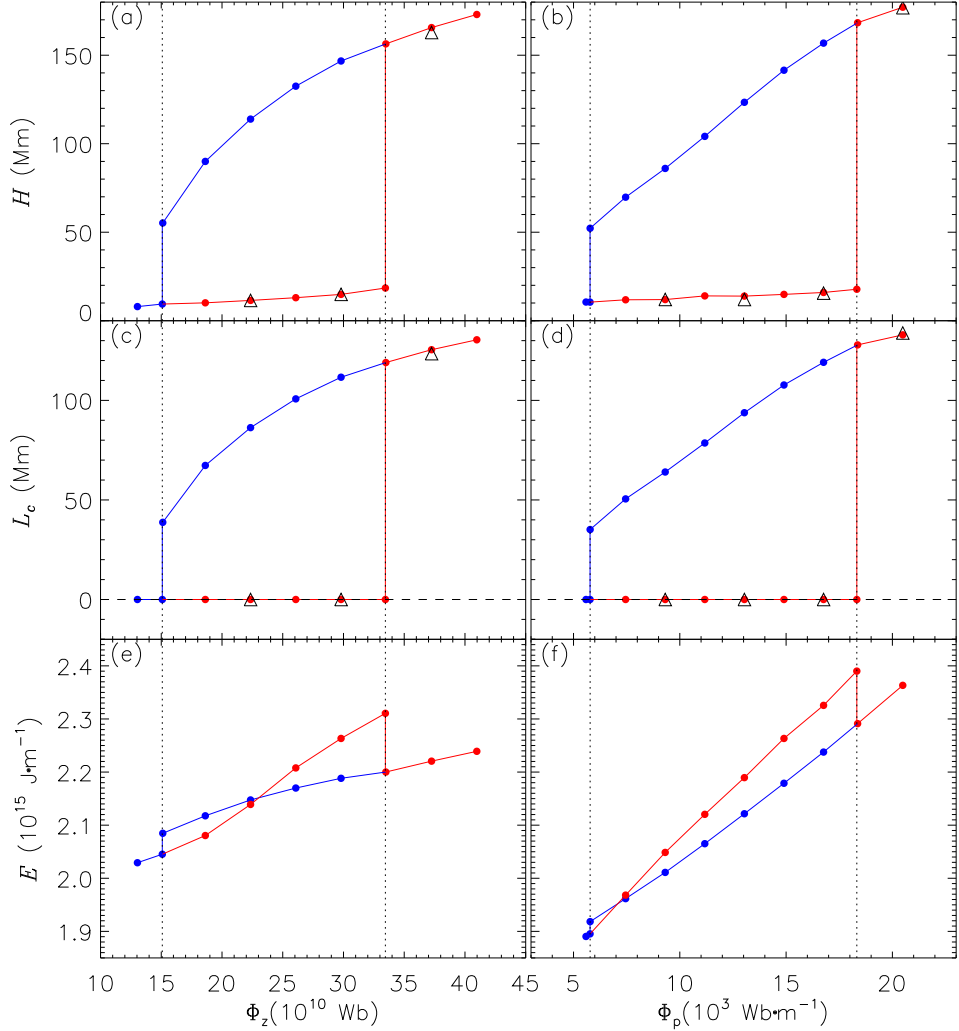}}
\caption{The height of the flux rope axis ($H$), the length of the current sheet below the rope ($L_c$), and the magnetic energy in the computational
domain ($E$) are shown as functions of the control parameter $\Phi_z$ for $\Phi_p = 14.9\times 10^3\mathrm{~Wb\cdot m^{-1}}$ in the left three panels, and as functions of the control parameter $\Phi_p$ for $\Phi_z = 29.8\times 10^{10}\mathrm{~Wb}$ in the right three panels. The red dots represent equilibrium solutions, which are obtained by increasing the related control parameter, proceeding from the lower branch to the upper via the upward catastrophic point, whereas the blue dots represent those, which are obtained by decreasing the related control parameter, proceeding from the upper branch to the lower via the downward catastrophic point. Finally, the solutions denoted by triangles are obtained from the product of the downward catastrophe and a monotonic increase of $\Phi_z$, and they follow almost the same profile as the red dots do.}\label{fig:pf}
\end{figure*}

First, in order to analyse the evolution of the equilibrium solutions versus $\Phi_z$, we focus on the solutions with different $\Phi_z$ but the same $\Phi_p=14.9\times10^{3}\mathrm{~Wb~m}^{-1}$. The geometric parameters describing these equilibrium solutions are plotted by the red dots in \fig{fig:pf}(a) and \ref{fig:pf}(c). The flux rope keeps sticking to the photosphere for $\Phi_z<33.5\times10^{10}\mathrm{~Wb}$ (see \fig{fig:conf}(a)) and the equilibrium solutions remain at the lower branch. Once the control parameter reaches $\Phi_z=33.5\times10^{10}\mathrm{~Wb}$, however, the flux rope jumps to the equilibrium state at the upper branch, levitating in the corona, as shown in \fig{fig:conf}(b). This indicates that a catastrophe with the flux rope jumping from the lower branch to the upper branch takes place, which is the upward catastrophe analysed in previous studies, and the critical value of the control parameter $\Phi_z$ is called the upward catastrophic point: $\Phi_z^u=33.5\times10^{10}\mathrm{~Wb}$. Our simulation reveals that the upward catastrophe also occurs in partially open bipolar field under force-free approximations.
\par
As mentioned above, all the previous studies only analysed the upward catastrophe. Here are the questions to be considered: What will happen to the flux rope at the upper branch with decreasing $\Phi_z$? Will the flux rope fall back to the lower branch? If so, will the flux rope fall back by a catastrophic jump or a continuous transition? To solve these problems, we start from the equilibrium solution at the upper branch obtained above (\fig{fig:conf}(b)) and decrease $\Phi_z$ to calculate the equilibrium solutions at the upper branch with similar procedures, i.e. from 0 to $100 \tau_A$ decrease $\Phi_z$ to a certain value and from $100 \tau_A$ to $200 \tau_A$ the system relaxes to equilibrium state with the given $\Phi_z$. $\Phi_p$ is still fixed to be $14.9\times10^{3}\mathrm{~Wb~m}^{-1}$. As shown by the blue dots in \fig{fig:pf}(a) and \ref{fig:pf}(c), with decreasing $\Phi_z$, the flux rope does not fall back to the lower branch at $\Phi_z^u$, indicating that the upward catastrophe is an irreversible process. The simulation result reveals that the flux rope also keeps suspended in the corona, and the associated equilibrium solutions remain at the upper branch, till $\Phi_z=15.0\times10^{10}\mathrm{~Wb}$ (\fig{fig:conf}(c)), at which the flux rope suddenly falls back to the photosphere (\fig{fig:conf}(d)). This indicates that there exists another kind of catastrophe, during which the flux rope falls down from the upper branch to the lower branch, i.e. a downward catastrophe. The downward catastrophic point is $\Phi_z^d=15.0\times10^{10}\mathrm{~Wb}$. Furthermore, we increase $\Phi_z$ again, but from the lower-branch solution right after the downward catastrophe (\fig{fig:conf}(d)), to calculate equilibrium solutions with varying $\Phi_z$, as shown by the black triangles in \fig{fig:pf}(a) and \ref{fig:pf}(c). Similarly, the flux rope does not jump to the upper branch at $\Phi_z^d$ as well, indicating that the downward catastrophe is also an irreversible process. From \fig{fig:pf}(a) and \ref{fig:pf}(c), it can be seen that the equilibrium solutions calculated from the solution right after the downward catastrophe (black triangles) follow almost the same profile as the red dots do. When $\Phi_z$ reaches the catastrophic point, an upward catastrophe takes place as before. Consequently, a circulation is formed by these two different kinds of catastrophes (see \sect{sec:profile}). The values of the catastrophic points are close to the observed magnetic fluxes of CMEs \citep[e.g.][]{Wang2015a}.
\par
There are also upward and downward catastrophes as the annular magnetic flux $\Phi_p$ is taken to be the control parameter with fixed $\Phi_z=29.8\times10^{10}\mathrm{~Wb}$. By similar procedures mentioned above, the evolution of equilibrium solutions versus $\Phi_p$ is obtained, as plotted in \fig{fig:pf}(b) and \ref{fig:pf}(d). The meanings of the symbols are the same: red dots denote the process of searching for the upward catastrophe, blue for the downward catastrophe, and black triangles are the equilibrium states calculated from the state after the downward catastrophe has taken place. Both upward and downward catastrophes exist and belong to irreversible processes. The upward catastrophic point is $\Phi_p^u=18.4\times10^{3}\mathrm{~Wb~m}^{-1}$, and the downward catastrophic point is $\Phi_p^d=5.78\times10^{3}\mathrm{~Wb~m}^{-1}$. 
\par

\subsection{Evolution of the magnetic energy}
\label{sec:energy}
As pointed in \sect{sec:catastrophe}, there exist upward and downward catastrophes; they are both irreversible processes, but the flux rope moves in opposite directions. Whether magnetic energy is released or stored during the downward catastrophe will help to understand the nature of these two catastrophes. Since the magnetic energy per unit length along the $z$-axis is infinite for a partially open bipolar field  in Cartesian coordinates, the evolution of magnetic energy has never been touched upon before. As an expedient measure, we evaluate the magnetic energy in the computational domain only, and think it approximately represents the evolutionary tendency of the magnetic energy of the whole system. As long as the domain is sufficiently large in size, as we did and explained in \sect{sec:equation}, the conclusions obtained should be qualitatively correct.
\par
The magnetic energy per unit length in $z$-direction in the computational domain is calculated by:
\begin{align}
E=\int\int\frac{B^2}{2\mu_0}dxdy.
\end{align}
The evolutions of the magnetic energy within the domain are plotted in \fig{fig:pf}(e) and \fig{fig:pf}(f) with $\Phi_z$ and $\Phi_p$ as the control parameter, respectively. Magnetic energy is released during both of the two catastrophes. It is natural that magnetic energy is always released during catastrophes: any catastrophe belongs to a spontaneous process in the absence of external influences, and the system must evolve to a state of lower energy after catastrophe. We may conclude that, although different kinds of catastrophes could have different kinematic characters (e.g. in this paper the moving directions of the flux rope are opposite during the upward and downward catastrophes), catastrophe should always correspond to a sudden energy release. By assuming that spatial scale in $z$-direction is the same as $x$-direction and $y$-direction, i.e. $\sim300\mathrm{~Mm}$, the released energy of the catastrophes is estimated to be: $3.3\times10^{22}$ J for the upward catastrophe and $1.2\times10^{22}$ J for the downward catastrophe with $\Phi_z$ as the control parameter; $3.0\times10^{22}$ J for the upward and $6.9\times10^{21}$ J for the downward with $\Phi_p$ as the control parameter. More magnetic energy is released for the upward catastrophe than the downward. It is revealed that the released magnetic energy during the catastrophes is of the order of $10^{22}$ J. It should be noted that the amount of the released energy is inversely proportional to the value of $\beta_0$, which is set to be 0.1 in this study. Larger amount of released magnetic energy can be obtained with smaller $\beta_0$. For example, if $\beta_0=0.001$, which is more proper for coronal situation \citep{Amari1999a}, the released energy should be as large as $\sim10^{24}$ J, which is comparable to the released energy of a medium flare. 
\par

\subsection{A cyclic evolution of the flux rope system}
\label{sec:profile}
\begin{figure*}
\includegraphics[width=\hsize]{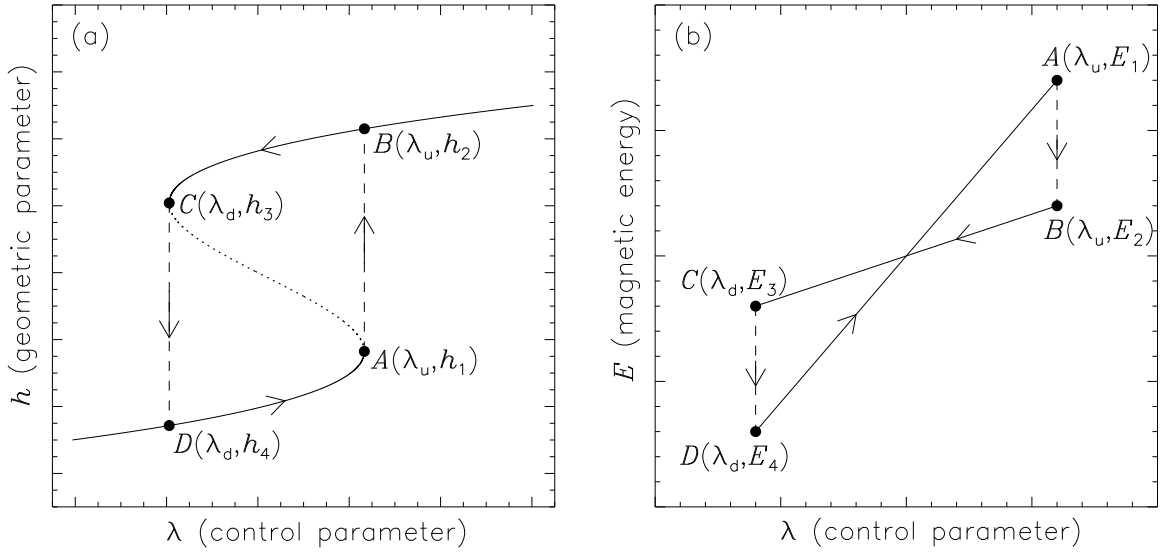}
\caption{Schematic representations of the equilibrium solutions of the magnetic flux rope system: the profile of (a) a certain geometric parameter of the rope ($h$) and (b) the magnetic energy in the computational domain ($E$) versus a certain control parameter of the system ($\lambda$). An upper catastrophe occurs at $\lambda = \lambda_u$, whereas a downward catastrophe occurs at $\lambda = \lambda_d$. The subscripts 1, 2, 3, and 4 label the state of the system: right before and after the upward catastrophic point, and right before and after the downward catastrophic point, respectively. Although the jump of $h$ is opposite in sign, the jump of $E$ is always negative, implying a magnetic energy release for both of the two catastrophes.}\label{fig:cart}
\end{figure*}

The evolution of a flux rope in a partially open bipolar background field as discussed in \sect{sec:catastrophe} and \sect{sec:energy} is summarised in \fig{fig:cart}, where $h$ stands for the geometric parameter and $\lambda$ the control parameter. There are two branches of equilibrium states for the flux rope: the lower branch ($D\sim A$ in \fig{fig:cart}) and the upper branch ($C\sim B$ in \fig{fig:cart}). The two branches are separated by the upward (from lower to upper branch) and downward (from upper to lower branch) catastrophes. The upward (downward) catastrophe is an irreversible but reproducible process, and it takes place when the solution jumps from the lower (upper) branch to the upper (lower) branch with increasing (decreasing) control parameter. Therefore, the two branches and the two catastrophes form a circulation, as shown by $D\rightarrow A\rightarrow B\rightarrow C\rightarrow D$ in \fig{fig:cart}(a). 
\par
The evolution of magnetic energy (\fig{fig:cart}(b)) is also a circulation consisting of the two branches and two catastrophes, but it is quite different from that of the geometric parameter (\fig{fig:cart}(a)). At the lower branch ($D\rightarrow A$), magnetic energy is stored with increasing control parameter, whereas it is reduced during both upward and downward catastrophes, and during the shift of the solution along the upper branch from $B$ to $C$ as well. The constraint exerted by the background field on the flux rope is larger for solutions at the lower branch than that at the upper branch, so that the growth rate of the magnetic energy with $\lambda$, $dE/d\lambda$, is larger for lower-branch solutions than upper-branch solutions. Consequently, at the upward catastrophic point ($\lambda=\lambda_u$), the magnetic energy of the upper-branch solution is smaller than that of the lower branch solution ($E_2<E_1$), whereas at the downward catastrophic point ($\lambda=\lambda_d$), the opposite is true ($E_3>E_4$). This provides the essential energy condition under which the downward catastrophe could occur.
\par

\subsection{Mechanism of magnetic energy release during catastrophe}
\label{sec:work}
It is widely accepted that magnetic energy is released by magnetic reconnection in solar eruptive activities \citep{Benz2008a,Shibata2011a}. However, magnetic reconnections are prohibited during our simulation (see \sect{sec:equation}). There should be other mechanisms of magnetic energy release in addition to magnetic reconnection. By simulations in spherical coordinates, \cite{Chen2006b} found that the magnetic force acting on the flux rope vanishes for equilibrium states, but becomes upward if the rope erupts after the upward catastrophe. As a result, the flux rope is expected to be accelerated by the Amp\`{e}re's force. Here, we calculate the work done by Amp\`{e}re's force during catastrophe in the entire computational domain, so as to reveal the mechanism of magnetic energy release.
\par
\begin{figure*}
\includegraphics[width=\hsize]{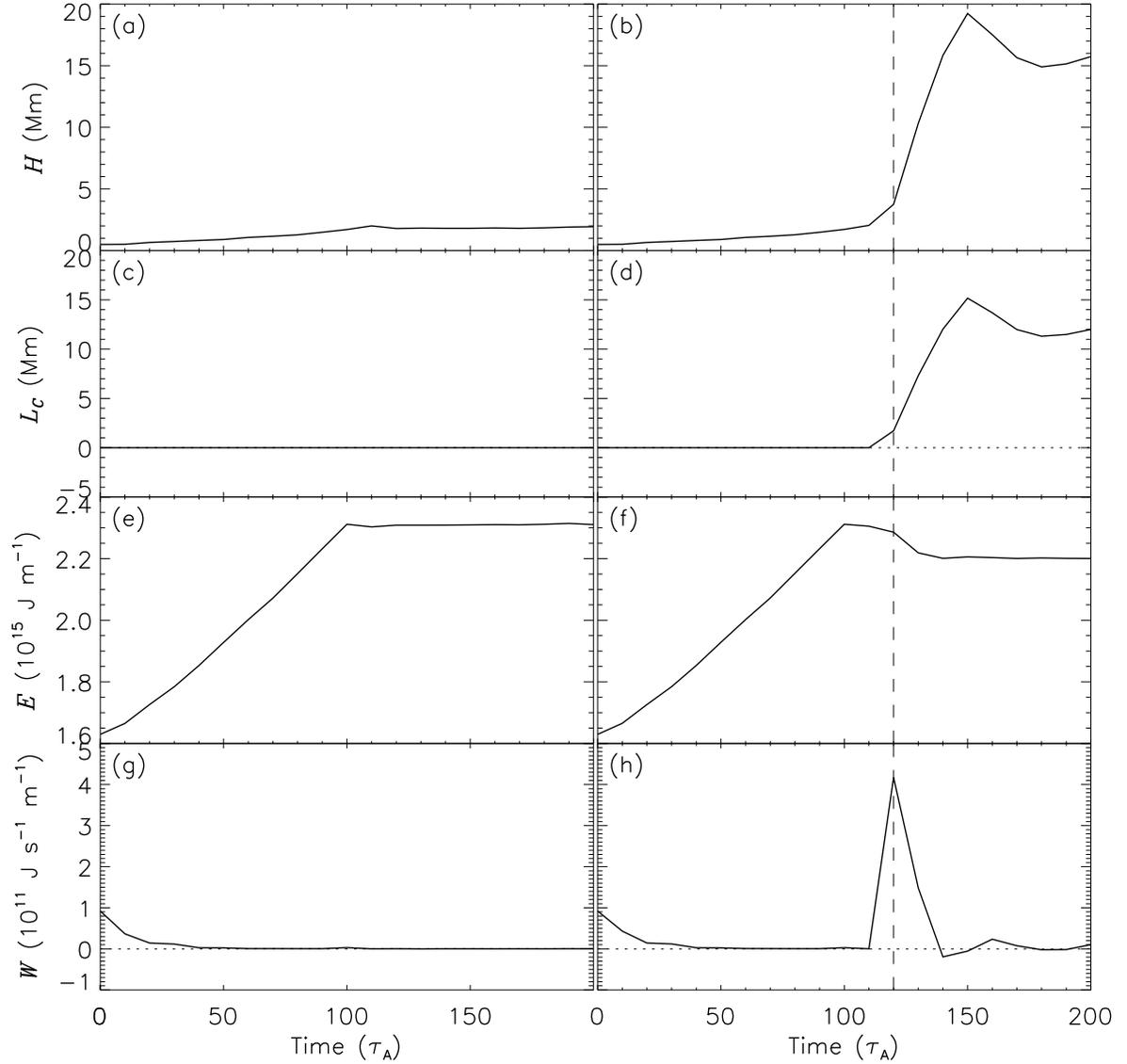}
\caption{The temporal evolution of the height of the flux rope axis ($H$), the length of the current sheet below the rope ($L_c$), the magnetic energy in the computational domain ($E$), and the rate of doing work by the Amp\`{e}re's force ($W$) during a transition from the initial state to the solution with a given annular flux of the rope, $\Phi_p = 14.9\times 10^3\mathrm{~Wb\cdot m^{-1}}$, are shown for the axial flux of the rope $\Phi_z = 33.4 \times 10^{10}\mathrm{~Wb}$
in the left four panels and $\Phi_z = 33.5 \times 10^{10}\mathrm{~Wb}$ in the right four panels. The axial flux changes gradually to the assigned value from $t=0$ to $100\tau_A$ and keeps invariant from $100\tau_A$ to $200\tau_A$, so as to reach a new equilibrium. The vertical dashed lines in the right panels denote the time at which the rate of doing work by Amp\`{e}re's force reaches its peak value.}\label{fig:up}
\end{figure*}

\fig{fig:up} shows the temporal profiles of relevant geometric and physical parameters during a transition toward the equilibrium solutions right before and after the upward catastrophe, which correspond to $(\Phi_z,\Phi_p)=(33.4\times10^{10}\mathrm{~Wb},14.9\times10^{3}\mathrm{~Wb~m}^{-1})$ and $(33.5\times10^{10}\mathrm{~Wb},14.9\times10^{3}\mathrm{~Wb~m}^{-1})$, respectively. The first four panels show the geometric parameters, including $H$ and $L_c$, whereas panels (e) and (f) show the magnetic energy. Finally, panels (g) and (h) depict the rate of doing work by Amp\`{e}re's force, which is determined by:
\begin{align}
&W=\int\int\overrightarrow{j}\cdot\overrightarrow{E}dxdy=\int\int-\overrightarrow{j}\cdot(\overrightarrow{v}\times\overrightarrow{B})dxdy= \int\int\left[\frac{1}{\mu_0}(\triangledown\times\overrightarrow{B})\times\overrightarrow{B}\right] \cdot\overrightarrow{v}dxdy,
\end{align}
where $\overrightarrow{j}$ and $\overrightarrow{E}$ denote the current density and electric field. 
\par
From 0 to $100 \tau_A$, $\Phi_z$ and $\Phi_p$ are smoothly adjusted. Note that at the beginning (0$\sim$40$\tau_A$), $W$ is small but finite. This should result from the deviation of the system from equilibrium at the beginning caused by adjusting the control parameters. From $100 \tau_A$ to $200 \tau_A$, during which $\Phi_z$ and $\Phi_p$ is constant, the magnetic energy is almost constant if the upward catastrophe does not take place (\fig{fig:up}(e)), and $W$ is almost zero accordingly (\fig{fig:up}(g)). If the upward catastrophe takes place, however, not only magnetic energy is released (\fig{fig:up}(f)), but also a significant peak value of $W$ appears in \fig{fig:up}(h). The total work done by Amp\`{e}re's force during upward catastrophe is roughly estimated to be about $1.0\times10^{14}\mathrm{~J~m^{-1}}$, in the same order of the released magnetic energy ($\sim 10^{14}\mathrm{~J~m^{-1}}$). The deviation might result from the Poynting energy flow and the influence of relaxation method. Therefore, we may conclude that the magnetic energy is released primarily by the work done by Amp\`{e}re's force during the upward catastrophe. For downward catastrophe, although the deviation of the work done by Amp\`{e}re's force from the released energy is larger, basic conclusion remains the same as for the upward catastrophe: the work done by Amp\`{e}re's force appears as a positive peak during the downward catastrophe.
\par

\section{Discussion and conclusion}
\label{sec:discuss}
To investigate the catastrophic behavior of coronal flux ropes, we simulate the evolution of the equilibrium states associated with a flux rope in a force-free partially open bipolar field versus different control parameters. It is found that, under force-free approximation, there is also upward catastrophe. What's more, under the force-free approximation, there exists another possibility that a downward catastrophe takes place, during which a levitating flux rope may fall down to the photosphere. The evolutionary scenario represented by the control parameter might cause a catastrophe to take place. For example, the ``flux-feeding'' procedure, during which chromospheric fibrils rise upward and merge with the prominence above \citep{Zhang2014a}, will result in varying $\Phi_z$, and varying $\Phi_p$ could be caused by twisting or untwisting motions of the flux rope \citep{Torok2010a,Liu2016a}. All these phenomena are the possible triggers of catastrophes. Upward and downward catastrophes connect the two branches of equilibrium states so as to form a cycle. These two catastrophes are both irreversible but reproducible processes. Thus there might exist activities during which more than one catastrophe take place: e.g. a flux rope is suspended in the corona at first, then falls back to the photosphere, and at last jumps upward, resulting in an eruptive activity.
\par
By calculating the magnetic energy within the numerical domain, the evolution of magnetic energy is analysed semi-quantitatively. Although the moving directions of the rope are opposite for the upward and downward catastrophes, magnetic energy is always released. The order of the released energy is rather large, comparable to a medium flare (see \sect{sec:energy}). Since there are no magnetic reconnections, magnetic energy is mainly released by the work done by Amp\`{e}re's force. Our calculation demonstrates that the magnetic energy released by the work done by Amp\`{e}re's force during catastrophe is sufficient for solar eruptive activities, indicating that magnetic reconnection is not always necessary. If magnetic reconnection is included in the simulation, the eruptive speed can be signiﬁcantly enhanced \citep{Chen2007a}, and magnetic energy is released by both magnetic reconnection and the work done by Amp\`{e}re's force.
\par
Previous studies have proposed that an upward catastrophe can serve as an effective mechanism for solar eruptive activities. Since catastrophe occurs via a loss of equilibrium at a critical point, it can be triggered by very small disturbances. In addition, upward catastrophes can not only account for CMEs and flares, they also provide sites for fast reconnection \citep{Chen2007a}. Apart from eruptive activities, there are also energetic but non-eruptive activities, such as confined flares \citep{Liu2014a}. The physical mechanism of confined flares has been discussed in many previous studies \citep[e.g.][]{Yang2014a,Joshi2015a}. During a downward catastrophe, although magnetic energy is still released, the flux rope falls back to the photosphere. Therefore, a downward catastrophe might be another possible cause of energetic but non-eruptive activities. Observational evidences are still needed to confirm these conjectures.

\par
This research is supported by Grants from NSFC 41131065, 41574165, 41421063 and 41222031, MOEC 20113402110001, CAS Key Research Program KZZD-EW-01-4, and the fundamental research funds for the central universities WK2080000077.

%\bibliographystyle{apj}
%\bibliography{catastrophe}

\end{document}